\documentclass[pre]{revtex4}
\usepackage{amsmath,amssymb,latexsym,epsfig,graphics,epsf,bm}
\usepackage{graphics}
\usepackage{float}
\usepackage{placeins}
\newcommand{\br}{\mathbf r}

\newcommand{\tbr}{\tilde{\mathbf r}}
\newcommand{\bF}{\mathbf F}
\newcommand{\tbF}{\tilde{\mathbf F}}
\newcommand{\bR}{\mathbf R}
\newcommand{\tbR}{\tilde{\mathbf R}}
\newcommand{\Ginf}{G_\infty}
\newcommand{\tti}{\xi}
\newcommand{\tphi}{{\phi}}
\newcommand{\beq}{\begin{equation}}
\newcommand{\eeq}{\end{equation}}

\newcommand{\Cvex}{C_V}%^{\rm ex}}
\newcommand{\cvex}{c_V}%^{\rm ex}}
\newcommand{\Sex}{S}%_{\rm ex}}
\newcommand{\sex}{s}%_{\rm ex}}
\newcommand{\Fex}{F_{\rm ex}}
\newcommand{\tU}{\tilde{\Phi}}
\renewcommand{\o}{\char 28}
\usepackage{color}
\begin{document}

\title{Direct tests of single-parameter aging}
\author{Tina Hecksher, Niels Boye Olsen, and Jeppe C. Dyre}\email{dyre@ruc.dk}
\affiliation{DNRF Center ``Glass and Time'', IMFUFA, Dept. of Sciences, Roskilde University, P. O. Box 260, DK-4000 Roskilde, Denmark}
\date{\today}

\begin{abstract}
This paper presents accurate data for the physical aging of organic glasses just below the glass transition probed by monitoring the following quantities after temperature up and down jumps: the shear-mechanical resonance frequency ($\sim$ 360 kHz), the dielectric loss at 1 Hz, the real part of the dielectric constant at 10 kHz, and the loss-peak frequency of the dielectric beta process ($\sim$ 10 kHz). The setup used allows for keeping temperature constant within 100 $\mu$K and for thermal equilibration within a few seconds after a temperature jump. The data conform to a new simplified version of the classical Tool-Narayanaswamy aging formalism, which makes it possible to calculate one relaxation curve directly from another without any fitting to analytical functions.
\end{abstract}

\maketitle

Gradual changes of material properties are referred to as aging. These are often caused by slow chemical reactions, but in some cases they reflect so-called {\it physical aging} that results exclusively from changes in atomic or molecular positions \cite{sim31,too31,kov63,nar71,moy76,moy76a,maz77,str78,kov79,scherer,hod94,ang00,whi06,koh13}. For applications it is important to be able to predict how fast material properties change over time, as well as in production \cite{sul90,hod95,hut95,ode11,can13,can14}. For instance, the performance of a smartphone's display glass is governed by the volume relaxation taking place when the glass is cooled through the glass transition \cite{mauro}. 

Physical aging has been studied in publications dealing with the aging of, e.g., oxide glasses \cite{nar71,moy76,moy76a,maz77}, polymers \cite{kov63,str78,hod95,hut95,gra12,can13}, metallic glasses \cite{che78,qia14}, spin glasses \cite{lun83,ber02}, relaxor ferroelectrics \cite{kir02}, and soft glassy materials like colloids and gels \cite{fie00,fof04}. Quantities probed to monitor aging are, e.g., density \cite{kov63,spi66}, enthalpy \cite{nar71,moy76}, Young's modulus \cite{che78}, gas permeability \cite{hua04}, high-frequency mechanical moduli \cite{ols98,dil04}, dc conductivity \cite{str78}, frequency-dependent dielectric constant \cite{sch91,leh98,lun05,ric15}, XPCS-probed structure \cite{rut12}, non-linear dielectric susceptibility \cite{bru12}, etc. 

Physical aging is generally nonexponential in time and nonlinear in temperature variation. Our focus below is on the aging of glasses just below their glass transition temperature, which is characterized by self-retardation for temperature down jumps and self-acceleration for up jumps \cite{too46,scherer,mck95,mau09a,can13}. The standard aging formalism is due to Narayanaswamy, an engineer at Ford Motor Company who back in 1970 needed a theory for predicting how the frozen-in stresses in a windshield depend on the glass' thermal history. The resulting so-called Tool-Narayanaswamy (TN) theory accounts for the nonexponential and nonlinear nature of aging, as well as the crossover (Kovacs) effect demonstrating memory of the thermal history \cite{scherer,rit56}. The TN trick is to assume the existence of an ``inner clock'' that defines a so-called material time \cite{hop58,mor60,lee63,mck94}. This is like the proper-time concept of the theory of relativity giving the time measured on a clock traveling with the observer. During aging the clock rate itself ages, which causes nonlinearity in temperature variation. A crucial assumption of the TN theory is that the ``fictive temperature'' controls both the clock rate and the quantity being monitored. This single-parameter assumption is usually tested by fitting data to analytical functions; below we develop a simplified TN theory that may be tested directly from data without any fitting.

This paper presents accurate temperature-jump aging data for organic glasses obtained by monitoring the following four quantities: the high-frequency shear-mechanical resonance frequency \cite{ols98}, the low-frequency dielectric loss (data from Ref. \cite{hec10}) \cite{sch91,weh07}, the high-frequency real part of the dielectric constant \cite{weh07}, and the dielectric loss-peak frequency of the beta process (data partly published in Ref. \cite{dyr03}). 
The setup used is described in Refs. \cite{iga08a,iga08b}. It is based on a custom-made cryostat capable of keeping temperature constant within 100$\mu$K for the first three quantities and within 1 mK for the fourth. A Peltier element is used for the cryostat's inner temperature control, and the time constant for equilibration of the setup after a temperature jump is only two seconds. The dielectric measurements were made with a homebuilt setup that uses a digital frequency generator below 100 Hz producing a sinusoidal signal with voltages reproducible within 10 ppm; at higher frequencies a standard LCR meter is used. The mechanical resonance measurements were carried out using a one-disc version of our piezo-ceramic shear transducer \cite{chr95}. More details are given in the Supplemental Material.

\begin{figure}
\begin{center}
\includegraphics[width=8cm]{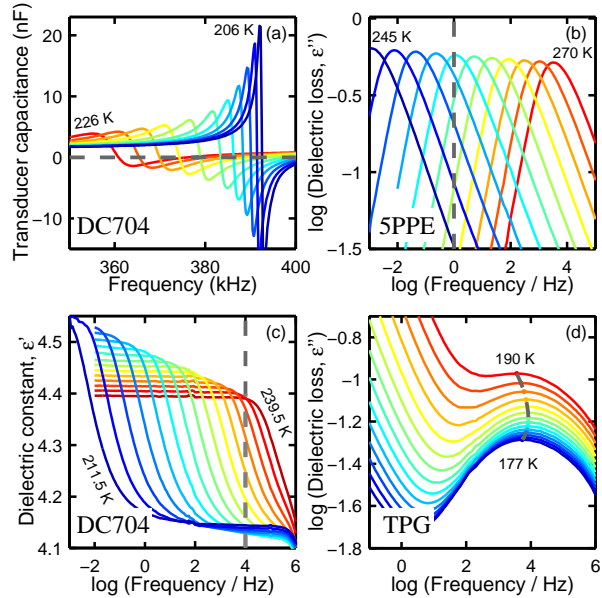}
\end{center}
\caption{\label{rawdata}Spectra illustrating the different types of measurements used for probing physical aging. The intersections between the grey dashed lines and the spectra indicate the quantities monitored. (a) Mechanical resonance of a piezo-electric disc; (b) dielectric loss; (c) real part of the dielectric constant at a high frequency, which at the aging temperatures is much above the alpha relaxation frequency; (d) dielectric beta relaxation loss-peak frequency (raw loss-peak frequency with no correction for the alpha process which, however, is insignificant at the aging temperatures). }
\end{figure}

The three liquids studied are tetramethyl-tetraphenyl-trisiloxane (DC704), 5-polyphenyl-4-ether (5PPE), and tripropylene glycol (TPG). Examples of the measurements behind the aging analysis are given in Fig. 1 (for a more thorough discussion please refer to the Supplemental Material). Figure 2 shows how the monitored quantity $X(t)$ equilibrates upon temperature up and down jumps (black and light blue). There is always a rapid change of $X$. The subsequent aging starts from a short-time plateau, which is most clearly visible for the up-jump data points.

\begin{figure}
\begin{center}
  \includegraphics[width=7cm]{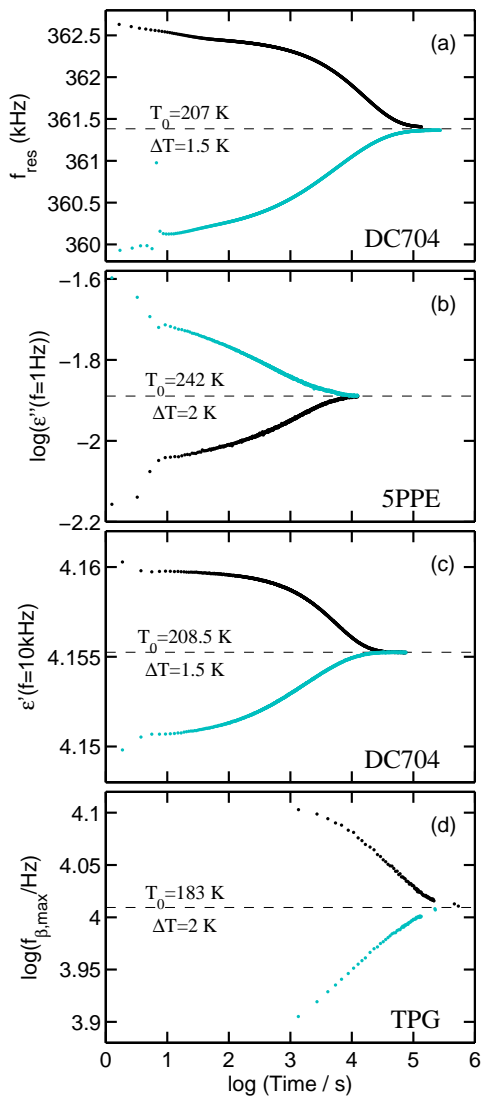}
\end{center}
\caption{\label{Xcurves}Aging data for temperature jumps for the different quantities probed. Each panel shows a pair of $\Delta T$ up and down jumps to the same temperature $T_0$. Up jumps are shown in black and have the characteristic self-accelerated shape, down jumps (in light blue) display self-retarded behavior. (a) Mechanical resonance frequency; (b) logarithm of the dielectric loss at 1~Hz; (c) real part of the dielectric constant at 10~kHz; (d) loss-peak frequency of the dielectric beta relaxation.}
\end{figure}

Consider a temperature jump initiated at $t=0$, which is studied by monitoring the time development of $X$. The jump starts from equilibrium at temperature $T_0+\Delta T$ and ends in equilibrium at $T_0$ at which the equilibrium value of $X$ is denoted by $X_{\rm eq}$. Following the convention of the aging literature the time-dependent variation of $X$ after the jump is denoted by $\Delta X(t)\equiv X(t)-X_{\rm eq}$. Thus $\Delta X(t)$ goes from $\Delta X(0)$ to zero as $t\rightarrow\infty$ and equilibrium at $T_0$ is attained.

The material time of the TN formalism denoted by $\tti$ is defined from the rate $\gamma(t)$ of the system's ``inner clock'' as follows

\beq\label{eq1}
d\tti=\gamma(t)\,dt\,.
\eeq
The TN formalism implies that for the general temperature variation $T_0+\Delta T(t)$ the quantity $\Delta X(t)$ can be written as an instantaneous contribution plus a material-time convolution integral \cite{nar71},

\beq\label{Nar}
\Delta X(\tti)=C\Delta T(\tti)\,-\,\int_{-\infty}^\tti M(\tti-\tti')\, \frac{d\Delta T}{d\tti'}(\tti')\,d\tti'\,.
\eeq
Here $\tti=\tti(t)$ found by integration of Eq. (\ref{eq1}). After a jump at $t=0$ from $T_0+\Delta T$ to $T_0$ it follows from Eq. (\ref{Nar}) that $\Delta X(t)=\Delta T (-C+M(\tti))$. 

We study jumps small enough that the jump magnitude obeys $\Delta X(0)\propto\Delta T$. In terms of the dimensionless function $\tphi(\tti)\equiv (dT/dX)(-C+M(\tti))$ one has $\Delta X(t)=\Delta X(0)\,\tphi(\tti)$ with $\tphi(0)=1$. Defining the normalized relaxation function $R(t)$ by  

\beq\label{Rt0}
R(t)\equiv\frac{\Delta X(t)}{\Delta X(0)}\,,
\eeq
for any temperature jump we thus have

\beq\label{Rt}
R(t)=\tphi(\tti)\,.
\eeq

We have so far followed Narayanaswamy's seminal 1971 paper \cite{nar71} and proceed to convert Eq. (\ref{Rt}) into a differential equation. Since $d\tti/dt=\gamma(t)$, the time derivative of $R$ is given by $\dot R=\tphi'(\tti)\gamma(t)$. Equation (\ref{Rt}) implies that $\tti$ is a unique function of $R$; thus $\tphi'(\tti)$ is also a unique function of $R$. Denoting this negative function by $-F(R)$ leads to

\beq\label{deq1}
\dot R \,=\, -\,F(R)\,\gamma(t)\,.
\eeq

Suppose a single parameter $Q$ controls both $X$ and the clock rate. The physical nature of $Q$ is irrelevant \cite{dyr03,ell07}. For small temperature jumps it is reasonable to assume that one can expand $X$ to first order in $Q$: $\Delta X\equiv X-X_{\rm eq}=c_1(Q-Q_{\rm eq})$ in which $Q_{\rm eq}$ is the equilibrium value of $Q$ at $T_0$ \cite{dyr03}. The clock rate is determined by barriers to be overcome and their activation energies, so one likewise expects a first-order expansion of the form $\ln\gamma-\ln\gamma_{\rm eq}=c_2(Q-Q_{\rm eq})$ to apply. Eliminating $Q-Q_{\rm eq}$ leads to $\ln\gamma=\ln\gamma_{\rm eq}+a\Delta X/X_{\rm eq}$ in which $a\equiv c_2X_{\rm eq}/c_1$ is a dimensionless constant. Introducing the time dependence explicitly via Eq. (\ref{Rt0}) we have \cite{too46,pet72}

\beq\label{gam}
\gamma(t)=\gamma_{\rm eq}\, \exp\left(a\frac{\Delta X(0)}{X_{\rm eq}}R(t)\right)\,.
\eeq
Substituting this into Eq. (\ref{deq1}) leads finally to the basic equation for single-parameter aging following a temperature jump,

\beq\label{deq2}
\dot R \,=\, -\,\gamma_{\rm eq}\,F(R)\, \exp\left(a\frac{\Delta X(0)}{X_{\rm eq}}R\right)\,.
\eeq

The important advance of Narayanaswamy in 1971 was to replace that time's {\it nonlinear} aging differential equations by a {\it linear} convolution integral. It may seem surprising that we now propose stepping back to a differential equation \cite{kol12}. Consistency with the TN formalism is ensured, however, by the fact that Eq. (\ref{deq2}) only applies for temperature jumps. In contrast, the aging differential equations of Tool and others of the form $d(X-X_{\rm eq}(T))/dt=-(X-X_{\rm eq}(T))/\tau(X,T)$ \cite{too46,rit56} were constructed to describe general temperature histories $T(t)$. Such equations lead to simple exponential relaxation in the linear aging limit ($\Delta T\rightarrow 0$), which is rarely observed, and they cannot account for the crossover effect \cite{scherer}.

\begin{figure}
\begin{center}
\includegraphics[width=6cm]{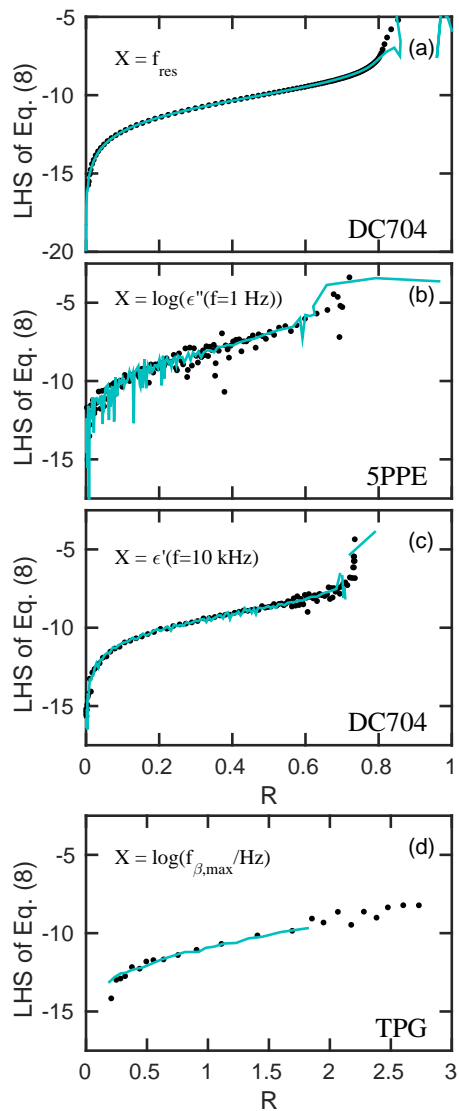}
\end{center}
\caption{\label{Pred1} Test of Eq. (\ref{Rrel}) for the Fig. \ref{Xcurves} data by plotting its left-hand side (except for the factor $\gamma_{\rm eq}$) as a function of the normalized relaxation function $R$ (time unit: seconds; the four $a$ parameters were determined from Eq. (\ref{tina_eq})). For the three first probes the instantaneous change after a temperature jump goes in the same direction as the subsequent aging. The beta loss-peak frequency initially jumps in the opposite direction, which is why $R$ is temporarily larger than unity for this data set (Fig. 4(d)). Data were binned and averaged over ten points, except in (d) where only five points were binned due to scarcity of data. As in Fig. 2 the black dots give the temperature up jumps; the down jumps are marked by a light blue curve connecting the small points.}
\end{figure}

Equation (\ref{deq2}) may be tested without fitting data to analytical functions or knowing $F(R)$. Taking the logarithm of Eq. (\ref{deq2}) leads to 

\beq\label{Rrel}
\ln\left(-\frac{\dot R}{\gamma_{\rm eq}}\right)-a\,\frac{\Delta X(0)}{X_{\rm eq}}\,R
\,=\,\ln\left(F(R)\right)\,.
\eeq
For any temperature jump the left-hand side is predicted to be a function of $R$ that is independent of the jump magnitude $\Delta X(0)$. This is tested in Fig. 3 by plotting the left-hand side against $R$ for the data of Fig. 2. The four $a$ parameters have not been optimized for the best fit; they were determined from Eq. (\ref{tina_eq}) derived below.

\begin{figure}
\begin{center}
\includegraphics[width=6cm]{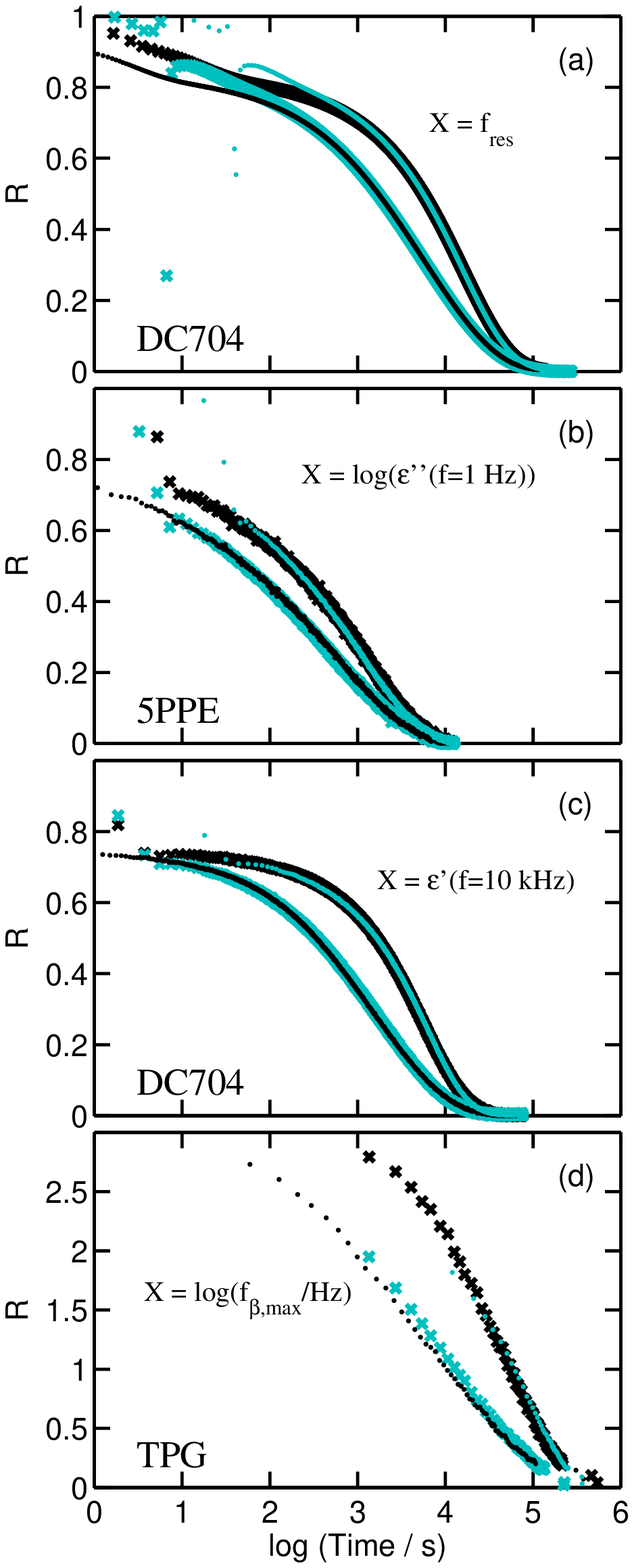}
\end{center}
\caption{\label{Pred2} Data (crosses) and predictions based on Eq. (\ref{tid2}) (dots) for each of the normalized Fig.~\ref{Xcurves} data sets. (a) Mechanical resonance frequency; (b) low-frequency dielectric loss; (c) high-frequency dielectric constant; (d) beta loss-peak frequency. Up jumps are in black, down jumps in light blue. Predicted curves are in the same color as the data they are calculated from. }
\end{figure}

A second test considers two temperature jumps to the same temperature $T_0$. The corresponding normalized relaxation functions are denoted by $R_1(t_1)$ and $R_2(t_2)$ with inverse functions $t_1(R_1)$ and $t_2(R_2)$. For times $t_1(R_1)$ and $t_2(R_2)$ corresponding to the same value of the normalized relaxation functions, $R_1=R_2$, Eq. (\ref{deq2}) implies

\beq\label{t1t2}
\frac{dR_1}{dt_1}\,\exp\left(-a\frac{\Delta X_1(0)}{X_{\rm eq}}R_1\right)
\,=\,\frac{dR_2}{dt_2}\,\exp\left(-a\frac{\Delta X_2(0)}{X_{\rm eq}}R_2\right)\,.
\eeq
For time increments $dt_1$ and $dt_2$ leading to identical changes $dR_1=dR_2$, if $\Lambda_{12}\equiv a(\Delta X_1(0)-\Delta X_2(0))/X_{\rm eq}$, Eq. (\ref{t1t2}) implies $dt_2=\exp({\Lambda_{12} R_1})dt_1$. By integration and identifying $R\equiv R_1=R_2$ this leads to

\beq\label{tid2}
t_2(R)
\,=\, \int_0^{t_2(R)} dt_2
\,=\,\int_0^{t_1(R)}e^{\Lambda_{12} R_1(t_1)}\,dt_1\,.
\eeq
This gives a simple recipe for calculating one normalized relaxation function from another. Figure 4 shows the normalized relaxation functions $R(t)$ of the Fig. 2 data (crosses) and those calculated from the other data set via Eq. (\ref{tid2}) (dots). 

Equation (\ref{tid2}) implies $t_2(R)-t_1(R)=\int_0^ {t_1(R)} (e^{\Lambda_{12} R_1(t_1)}-1) dt_1$. A similar expression applies for $t_1(R)-t_2(R)$. Since $\Lambda_{21}=-\Lambda_{12}$, adding the long-time limits of these expressions leads to the following consistency requirement

\beq\label{tina_eq}
\int_0^\infty\big(e^{\Lambda_{12} R_1(t_1)}-1\big)\,dt_1\,+\, \int_0^\infty\big(e^{-\Lambda_{12} R_2(t_2)}-1\big)\,dt_2   \,=0\,.
\eeq
Since $\Lambda_{12}$ determines $a$, this provides an equation for the $a$ values used in Figs. 3 and 4. The Supplemental Material shows that the $a$ parameters derived in this way are consistent with extrapolations from higher-temperature equilibrium measurements.

In summary, we have presented accurate data for temperature jumps of organic glasses and derived a simplified version of the Narayanswamy's 1971 aging theory that allows for direct data tests. The new tests do not involve any fitting to analytical functions, as is usually done. In Ref. \onlinecite{hec10} we also proposed a test of the Narayanaswamy theory not involving such fits, but it was more complicated than the present procedure and did not make predictions for how to calculate all temperare jumps from knowledge of a single one. The test of Ref. \onlinecite{hec10} involved calculating derivates numerically from data, which is an inherently noisy process. -- Crucially, Eq. (\ref{deq2}) involves {\it both} the normalized and the unnormalized relaxation functions, $R(t)$ and $\Delta X(t)=\Delta X(0)\,R(t)$. This is necessary because a differential equation for only $R(t)$ cannot account for the nonlinearity, whereas a differential equation involving only $\Delta X(t)$ cannot lead to nonexponentiality in the linear limit.

There are other approaches to describing physical aging than the standard TN theory \cite{lun05,kol12}. The common ``single-parameter'' assumption of all simple theories is that the quantity monitored correlates to the clock rate $\gamma$. This is also the main ingredient in the approach of Lunkenheimer {\it et al.}, which assumes a stretched-exponential aging function with a characteristic inverse relaxation time that itself ages according to the same stretched exponential \cite{lun05,ric13}. -- In continuing work we are currently investigating the possibility of using higher-order expansions for describing temperature jumps larger than those reported here.

\begin{acknowledgments}
Kristine Niss is thanked for several useful discussions. The center for viscous liquid dynamics ``Glass and Time'' is sponsored by the Danish National Research Foundation via grant DNRF61.
\end{acknowledgments}


\begin{thebibliography}{57}
\expandafter\ifx\csname natexlab\endcsname\relax\def\natexlab#1{#1}\fi
\expandafter\ifx\csname bibnamefont\endcsname\relax
  \def\bibnamefont#1{#1}\fi
\expandafter\ifx\csname bibfnamefont\endcsname\relax
  \def\bibfnamefont#1{#1}\fi
\expandafter\ifx\csname citenamefont\endcsname\relax
  \def\citenamefont#1{#1}\fi
\expandafter\ifx\csname url\endcsname\relax
  \def\url#1{\texttt{#1}}\fi
\expandafter\ifx\csname urlprefix\endcsname\relax\def\urlprefix{URL }\fi
\providecommand{\bibinfo}[2]{#2}
\providecommand{\eprint}[2][]{\url{#2}}

\bibitem[{\citenamefont{Simon.}(1931)}]{sim31}
\bibinfo{author}{\bibfnamefont{F.}~\bibnamefont{Simon}}, \bibinfo{journal}{Z.
  Anorg. Allg. Chem.} \textbf{\bibinfo{volume}{203}}, \bibinfo{pages}{219}
  (\bibinfo{year}{1931}).

\bibitem[{\citenamefont{Tool and Eichlin}(1931)}]{too31}
\bibinfo{author}{\bibfnamefont{A.~Q.} \bibnamefont{Tool}} \bibnamefont{and}
  \bibinfo{author}{\bibfnamefont{C.~G.} \bibnamefont{Eichlin}},
  \bibinfo{journal}{J. Am. Ceram. Soc.} \textbf{\bibinfo{volume}{14}},
  \bibinfo{pages}{276} (\bibinfo{year}{1931}).

\bibitem[{\citenamefont{Kovacs}(1963)}]{kov63}
\bibinfo{author}{\bibfnamefont{A.~J.} \bibnamefont{Kovacs}},
  \bibinfo{journal}{Fortschr. Hochpolym.-Forsch.} \textbf{\bibinfo{volume}{3}},
  \bibinfo{pages}{394} (\bibinfo{year}{1963}).

\bibitem[{\citenamefont{Narayanaswamy}(1971)}]{nar71}
\bibinfo{author}{\bibfnamefont{O.~S.} \bibnamefont{Narayanaswamy}},
  \bibinfo{journal}{J. Am. Ceram. Soc.} \textbf{\bibinfo{volume}{54}},
  \bibinfo{pages}{491} (\bibinfo{year}{1971}).

\bibitem[{\citenamefont{Moynihan
  et~al.}(1976{\natexlab{a}})\citenamefont{Moynihan, Easteal, DeBolt, and
  Tucker}}]{moy76}
\bibinfo{author}{\bibfnamefont{C.~T.} \bibnamefont{Moynihan}},
  \bibinfo{author}{\bibfnamefont{A.~J.} \bibnamefont{Easteal}},
  \bibinfo{author}{\bibfnamefont{M.~A.} \bibnamefont{DeBolt}},
  \bibnamefont{and} \bibinfo{author}{\bibfnamefont{J.}~\bibnamefont{Tucker}},
  \bibinfo{journal}{J. Am. Ceram. Soc.} \textbf{\bibinfo{volume}{59}},
  \bibinfo{pages}{12} (\bibinfo{year}{1976}{\natexlab{a}}).

\bibitem[{\citenamefont{Moynihan
  et~al.}(1976{\natexlab{b}})\citenamefont{Moynihan, Macedo, Montrose, Gupta,
  DeBolt, Dill, Dom, Drake, Easteal, Elterman et~al.}}]{moy76a}
\bibinfo{author}{\bibfnamefont{C.~T.} \bibnamefont{Moynihan}},
  \bibinfo{author}{\bibfnamefont{P.~B.} \bibnamefont{Macedo}},
  \bibinfo{author}{\bibfnamefont{C.~J.} \bibnamefont{Montrose}},
  \bibinfo{author}{\bibfnamefont{P.~K.} \bibnamefont{Gupta}},
  \bibinfo{author}{\bibfnamefont{M.~A.} \bibnamefont{DeBolt}},
  \bibinfo{author}{\bibfnamefont{J.~F.} \bibnamefont{Dill}},
  \bibinfo{author}{\bibfnamefont{B.~E.} \bibnamefont{Dom}},
  \bibinfo{author}{\bibfnamefont{P.~W.} \bibnamefont{Drake}},
  \bibinfo{author}{\bibfnamefont{A.~J.} \bibnamefont{Easteal}},
  \bibinfo{author}{\bibfnamefont{P.~B.} \bibnamefont{Elterman}},
  \bibnamefont{et~al.}, \bibinfo{journal}{Ann. NY Acad. Sci.}
  \textbf{\bibinfo{volume}{279}}, \bibinfo{pages}{15}
  (\bibinfo{year}{1976}{\natexlab{b}}).

\bibitem[{\citenamefont{{Mazurin}}(1977)}]{maz77}
\bibinfo{author}{\bibfnamefont{O.}~\bibnamefont{{Mazurin}}},
  \bibinfo{journal}{J. Non-Cryst. Solids} \textbf{\bibinfo{volume}{25}},
  \bibinfo{pages}{129} (\bibinfo{year}{1977}).

\bibitem[{\citenamefont{Struik}(1978)}]{str78}
\bibinfo{author}{\bibfnamefont{L.~C.~E.} \bibnamefont{Struik}},
  \emph{\bibinfo{title}{{Physical Aging in Amorphous Polymers and Other
  Materials}}} (\bibinfo{publisher}{Elsevier, Amsterdam},
  \bibinfo{year}{1978}).

\bibitem[{\citenamefont{Kovacs et~al.}(1979)\citenamefont{Kovacs, Aklonis,
  Hutchinson, and Ramos}}]{kov79}
\bibinfo{author}{\bibfnamefont{A.~J.} \bibnamefont{Kovacs}},
  \bibinfo{author}{\bibfnamefont{J.~J.} \bibnamefont{Aklonis}},
  \bibinfo{author}{\bibfnamefont{J.~M.} \bibnamefont{Hutchinson}},
  \bibnamefont{and} \bibinfo{author}{\bibfnamefont{A.~R.} \bibnamefont{Ramos}},
  \bibinfo{journal}{J. Polym. Sci. Polym. Phys.} \textbf{\bibinfo{volume}{17}},
  \bibinfo{pages}{1097} (\bibinfo{year}{1979}).

\bibitem[{\citenamefont{Scherer}(1986)}]{scherer}
\bibinfo{author}{\bibfnamefont{G.~W.} \bibnamefont{Scherer}},
  \emph{\bibinfo{title}{{Relaxation in Glass and Composites}}}
  (\bibinfo{publisher}{Wiley, New York}, \bibinfo{year}{1986}).

\bibitem[{\citenamefont{Hodge}(1994)}]{hod94}
\bibinfo{author}{\bibfnamefont{I.~M.} \bibnamefont{Hodge}},
  \bibinfo{journal}{J. Non-Cryst. Solids} \textbf{\bibinfo{volume}{169}},
  \bibinfo{pages}{211} (\bibinfo{year}{1994}).

\bibitem[{\citenamefont{Angell et~al.}(2000)\citenamefont{Angell, Ngai,
  McKenna, McMillan, and Martin}}]{ang00}
\bibinfo{author}{\bibfnamefont{C.~A.} \bibnamefont{Angell}},
  \bibinfo{author}{\bibfnamefont{K.~L.} \bibnamefont{Ngai}},
  \bibinfo{author}{\bibfnamefont{G.~B.} \bibnamefont{McKenna}},
  \bibinfo{author}{\bibfnamefont{P.~F.} \bibnamefont{McMillan}},
  \bibnamefont{and} \bibinfo{author}{\bibfnamefont{S.~W.}
  \bibnamefont{Martin}}, \bibinfo{journal}{J. Appl. Phys.}
  \textbf{\bibinfo{volume}{88}}, \bibinfo{pages}{3113} (\bibinfo{year}{2000}).

\bibitem[{\citenamefont{White}(2006)}]{whi06}
\bibinfo{author}{\bibfnamefont{J.~R.} \bibnamefont{White}},
  \bibinfo{journal}{Comptes Rendus Chimie} \textbf{\bibinfo{volume}{9}},
  \bibinfo{pages}{1396} (\bibinfo{year}{2006}).

\bibitem[{\citenamefont{Koh and Simon}(2013)}]{koh13}
\bibinfo{author}{\bibfnamefont{Y.~P.} \bibnamefont{Koh}} \bibnamefont{and}
  \bibinfo{author}{\bibfnamefont{S.~L.} \bibnamefont{Simon}},
  \bibinfo{journal}{Macromolecules} \textbf{\bibinfo{volume}{46}},
  \bibinfo{pages}{5815} (\bibinfo{year}{2013}).

\bibitem[{\citenamefont{Sullivan}(1990)}]{sul90}
\bibinfo{author}{\bibfnamefont{J.~L.} \bibnamefont{Sullivan}},
  \bibinfo{journal}{{Composites Science and Technology}}
  \textbf{\bibinfo{volume}{39}}, \bibinfo{pages}{207} (\bibinfo{year}{1990}).

\bibitem[{\citenamefont{Hodge}(1995)}]{hod95}
\bibinfo{author}{\bibfnamefont{I.~M.} \bibnamefont{Hodge}},
  \bibinfo{journal}{Science} \textbf{\bibinfo{volume}{267}},
  \bibinfo{pages}{1945} (\bibinfo{year}{1995}).

\bibitem[{\citenamefont{Hutchinson}(1995)}]{hut95}
\bibinfo{author}{\bibfnamefont{J.~M.} \bibnamefont{Hutchinson}},
  \bibinfo{journal}{Prog. Polym. Sci.} \textbf{\bibinfo{volume}{20}},
  \bibinfo{pages}{703} (\bibinfo{year}{1995}).

\bibitem[{\citenamefont{Odegard and Bandyopadhyay}(2011)}]{ode11}
\bibinfo{author}{\bibfnamefont{G.~M.} \bibnamefont{Odegard}} \bibnamefont{and}
  \bibinfo{author}{\bibfnamefont{A.}~\bibnamefont{Bandyopadhyay}},
  \bibinfo{journal}{J. Polym. Sci. Part B: Polym. Phys.}
  \textbf{\bibinfo{volume}{49}}, \bibinfo{pages}{1695} (\bibinfo{year}{2011}).

\bibitem[{\citenamefont{Cangialosi et~al.}(2013)\citenamefont{Cangialosi,
  Boucher, Alegria, and Colmenero}}]{can13}
\bibinfo{author}{\bibfnamefont{D.}~\bibnamefont{Cangialosi}},
  \bibinfo{author}{\bibfnamefont{V.~M.} \bibnamefont{Boucher}},
  \bibinfo{author}{\bibfnamefont{A.}~\bibnamefont{Alegria}}, \bibnamefont{and}
  \bibinfo{author}{\bibfnamefont{J.}~\bibnamefont{Colmenero}},
  \bibinfo{journal}{Soft Matter} \textbf{\bibinfo{volume}{9}},
  \bibinfo{pages}{8619} (\bibinfo{year}{2013}).

\bibitem[{\citenamefont{Cangialosi}(2014)}]{can14}
\bibinfo{author}{\bibfnamefont{D.}~\bibnamefont{Cangialosi}},
  \bibinfo{journal}{J. Phys.: Condens. Matter} \textbf{\bibinfo{volume}{26}},
  \bibinfo{pages}{153101} (\bibinfo{year}{2014}).

\bibitem[{\citenamefont{Mauro}(2015)}]{mauro}
\bibinfo{author}{\bibfnamefont{J.~C.} \bibnamefont{Mauro}},
  \emph{\bibinfo{title}{{Presentation at ``Unifying Concepts in Glass Physics.
  VI. Aspen, CO'' }}} (\bibinfo{year}{2015}).

\bibitem[{\citenamefont{Grassia and Simon}({2012})}]{gra12}
\bibinfo{author}{\bibfnamefont{L.}~\bibnamefont{Grassia}} \bibnamefont{and}
  \bibinfo{author}{\bibfnamefont{S.~L.} \bibnamefont{Simon}},
  \bibinfo{journal}{Polymer} \textbf{\bibinfo{volume}{53}},
  \bibinfo{pages}{3613} (\bibinfo{year}{{2012}}).

\bibitem[{\citenamefont{Chen}(1978)}]{che78}
\bibinfo{author}{\bibfnamefont{H.~S.} \bibnamefont{Chen}}, \bibinfo{journal}{J.
  Appl. Phys.} \textbf{\bibinfo{volume}{49}}, \bibinfo{pages}{3289}
  (\bibinfo{year}{1978}).

\bibitem[{\citenamefont{Qiao and Pelletier}(2014)}]{qia14}
\bibinfo{author}{\bibfnamefont{J.~C.} \bibnamefont{Qiao}} \bibnamefont{and}
  \bibinfo{author}{\bibfnamefont{J.~M.} \bibnamefont{Pelletier}},
  \bibinfo{journal}{J. Mater. Sci. Technol.} \textbf{\bibinfo{volume}{30}},
  \bibinfo{pages}{523} (\bibinfo{year}{2014}).

\bibitem[{\citenamefont{Lundgren et~al.}(1983)\citenamefont{Lundgren,
  Svedlindh, Nordblad, and Beckman}}]{lun83}
\bibinfo{author}{\bibfnamefont{L.}~\bibnamefont{Lundgren}},
  \bibinfo{author}{\bibfnamefont{P.}~\bibnamefont{Svedlindh}},
  \bibinfo{author}{\bibfnamefont{P.}~\bibnamefont{Nordblad}}, \bibnamefont{and}
  \bibinfo{author}{\bibfnamefont{O.}~\bibnamefont{Beckman}},
  \bibinfo{journal}{Phys. Rev. Lett.} \textbf{\bibinfo{volume}{51}},
  \bibinfo{pages}{911} (\bibinfo{year}{1983}).

\bibitem[{\citenamefont{Berthier and Bouchaud}(2002)}]{ber02}
\bibinfo{author}{\bibfnamefont{L.}~\bibnamefont{Berthier}} \bibnamefont{and}
  \bibinfo{author}{\bibfnamefont{J.-P.} \bibnamefont{Bouchaud}},
  \bibinfo{journal}{Phys. Rev. B} \textbf{\bibinfo{volume}{66}},
  \bibinfo{pages}{054404} (\bibinfo{year}{2002}).

\bibitem[{\citenamefont{Kircher and B{\"o}hmer}(2002)}]{kir02}
\bibinfo{author}{\bibfnamefont{O.}~\bibnamefont{Kircher}} \bibnamefont{and}
  \bibinfo{author}{\bibfnamefont{R.}~\bibnamefont{B{\"o}hmer}},
  \bibinfo{journal}{Eur. Phys. J. B} \textbf{\bibinfo{volume}{26}},
  \bibinfo{pages}{329} (\bibinfo{year}{2002}).

\bibitem[{\citenamefont{Fielding et~al.}(2000)\citenamefont{Fielding, Sollich,
  and Cates}}]{fie00}
\bibinfo{author}{\bibfnamefont{S.~M.} \bibnamefont{Fielding}},
  \bibinfo{author}{\bibfnamefont{P.}~\bibnamefont{Sollich}}, \bibnamefont{and}
  \bibinfo{author}{\bibfnamefont{M.~E.} \bibnamefont{Cates}},
  \bibinfo{journal}{J. Rheol.} \textbf{\bibinfo{volume}{44}},
  \bibinfo{pages}{323} (\bibinfo{year}{2000}).

\bibitem[{\citenamefont{Foffi et~al.}(2004)\citenamefont{Foffi, Zaccarelli,
  Buldyrev, Sciortino, and Tartaglia}}]{fof04}
\bibinfo{author}{\bibfnamefont{G.}~\bibnamefont{Foffi}},
  \bibinfo{author}{\bibfnamefont{E.}~\bibnamefont{Zaccarelli}},
  \bibinfo{author}{\bibfnamefont{S.}~\bibnamefont{Buldyrev}},
  \bibinfo{author}{\bibfnamefont{F.}~\bibnamefont{Sciortino}},
  \bibnamefont{and}
  \bibinfo{author}{\bibfnamefont{P.}~\bibnamefont{Tartaglia}},
  \bibinfo{journal}{J. Chem. Phys.} \textbf{\bibinfo{volume}{120}},
  \bibinfo{pages}{8824} (\bibinfo{year}{2004}).

\bibitem[{\citenamefont{Spinner and Napolitano}(1966)}]{spi66}
\bibinfo{author}{\bibfnamefont{S.}~\bibnamefont{Spinner}} \bibnamefont{and}
  \bibinfo{author}{\bibfnamefont{A.}~\bibnamefont{Napolitano}},
  \bibinfo{journal}{J. Res. NBS} \textbf{\bibinfo{volume}{70A}},
  \bibinfo{pages}{147} (\bibinfo{year}{1966}).

\bibitem[{\citenamefont{Huang and Paul}(2004)}]{hua04}
\bibinfo{author}{\bibfnamefont{Y.}~\bibnamefont{Huang}} \bibnamefont{and}
  \bibinfo{author}{\bibfnamefont{D.}~\bibnamefont{Paul}},
  \bibinfo{journal}{Polymer} \textbf{\bibinfo{volume}{45}},
  \bibinfo{pages}{8377 } (\bibinfo{year}{2004}).

\bibitem[{\citenamefont{Olsen et~al.}(1998)\citenamefont{Olsen, Dyre, and
  Christensen}}]{ols98}
\bibinfo{author}{\bibfnamefont{N.~B.} \bibnamefont{Olsen}},
  \bibinfo{author}{\bibfnamefont{J.~C.} \bibnamefont{Dyre}}, \bibnamefont{and}
  \bibinfo{author}{\bibfnamefont{T.}~\bibnamefont{Christensen}},
  \bibinfo{journal}{Phys. Rev. Lett.} \textbf{\bibinfo{volume}{81}},
  \bibinfo{pages}{1031} (\bibinfo{year}{1998}).

\bibitem[{\citenamefont{Di~Leonardo et~al.}(2004)\citenamefont{Di~Leonardo,
  Scopigno, Ruocco, and Buontempo}}]{dil04}
\bibinfo{author}{\bibfnamefont{R.}~\bibnamefont{Di~Leonardo}},
  \bibinfo{author}{\bibfnamefont{T.}~\bibnamefont{Scopigno}},
  \bibinfo{author}{\bibfnamefont{G.}~\bibnamefont{Ruocco}}, \bibnamefont{and}
  \bibinfo{author}{\bibfnamefont{U.}~\bibnamefont{Buontempo}},
  \bibinfo{journal}{Rev. Sci. Instrum.} \textbf{\bibinfo{volume}{75}},
  \bibinfo{pages}{2631} (\bibinfo{year}{2004}).

\bibitem[{\citenamefont{Schlosser and Sch{\"o}nhals}(1991)}]{sch91}
\bibinfo{author}{\bibfnamefont{E.}~\bibnamefont{Schlosser}} \bibnamefont{and}
  \bibinfo{author}{\bibfnamefont{A.}~\bibnamefont{Sch{\"o}nhals}},
  \bibinfo{journal}{Polymer} \textbf{\bibinfo{volume}{32}},
  \bibinfo{pages}{2135} (\bibinfo{year}{1991}).

\bibitem[{\citenamefont{Leheny and Nagel}(1998)}]{leh98}
\bibinfo{author}{\bibfnamefont{R.~L.} \bibnamefont{Leheny}} \bibnamefont{and}
  \bibinfo{author}{\bibfnamefont{S.~R.} \bibnamefont{Nagel}},
  \bibinfo{journal}{Phys. Rev. B} \textbf{\bibinfo{volume}{57}},
  \bibinfo{pages}{5154} (\bibinfo{year}{1998}).

\bibitem[{\citenamefont{Lunkenheimer et~al.}(2005)\citenamefont{Lunkenheimer,
  Wehn, Schneider, and Loidl}}]{lun05}
\bibinfo{author}{\bibfnamefont{P.}~\bibnamefont{Lunkenheimer}},
  \bibinfo{author}{\bibfnamefont{R.}~\bibnamefont{Wehn}},
  \bibinfo{author}{\bibfnamefont{U.}~\bibnamefont{Schneider}},
  \bibnamefont{and} \bibinfo{author}{\bibfnamefont{A.}~\bibnamefont{Loidl}},
  \bibinfo{journal}{Phys. Rev. Lett.} \textbf{\bibinfo{volume}{95}},
  \bibinfo{pages}{055702} (\bibinfo{year}{2005}).

\bibitem[{\citenamefont{Richert}(2015)}]{ric15}
\bibinfo{author}{\bibfnamefont{R.}~\bibnamefont{Richert}},
  \bibinfo{journal}{Adv. Chem. Phys.} \textbf{\bibinfo{volume}{156}},
  \bibinfo{pages}{101} (\bibinfo{year}{2015}).

\bibitem[{\citenamefont{Ruta et~al.}(2012)\citenamefont{Ruta, Chushkin, Monaco,
  Cipelletti, Pineda, Bruna, Giordano, and Gonzalez-Silveira}}]{rut12}
\bibinfo{author}{\bibfnamefont{B.}~\bibnamefont{Ruta}},
  \bibinfo{author}{\bibfnamefont{Y.}~\bibnamefont{Chushkin}},
  \bibinfo{author}{\bibfnamefont{G.}~\bibnamefont{Monaco}},
  \bibinfo{author}{\bibfnamefont{L.}~\bibnamefont{Cipelletti}},
  \bibinfo{author}{\bibfnamefont{E.}~\bibnamefont{Pineda}},
  \bibinfo{author}{\bibfnamefont{P.}~\bibnamefont{Bruna}},
  \bibinfo{author}{\bibfnamefont{V.~M.} \bibnamefont{Giordano}},
  \bibnamefont{and}
  \bibinfo{author}{\bibfnamefont{M.}~\bibnamefont{Gonzalez-Silveira}},
  \bibinfo{journal}{Phys. Rev. Lett.} \textbf{\bibinfo{volume}{109}},
  \bibinfo{pages}{165701} (\bibinfo{year}{2012}).

\bibitem[{\citenamefont{Brun et~al.}(2012)\citenamefont{Brun, Ladieu, L'Hote,
  Biroli, and Bouchaud}}]{bru12}
\bibinfo{author}{\bibfnamefont{C.}~\bibnamefont{Brun}},
  \bibinfo{author}{\bibfnamefont{F.}~\bibnamefont{Ladieu}},
  \bibinfo{author}{\bibfnamefont{D.}~\bibnamefont{L'Hote}},
  \bibinfo{author}{\bibfnamefont{G.}~\bibnamefont{Biroli}}, \bibnamefont{and}
  \bibinfo{author}{\bibfnamefont{J.-P.} \bibnamefont{Bouchaud}},
  \bibinfo{journal}{Phys. Rev. Lett.} \textbf{\bibinfo{volume}{109}},
  \bibinfo{pages}{175702} (\bibinfo{year}{2012}).

\bibitem[{\citenamefont{Tool}(1946)}]{too46}
\bibinfo{author}{\bibfnamefont{A.~Q.} \bibnamefont{Tool}}, \bibinfo{journal}{J.
  Am. Ceram. Soc.} \textbf{\bibinfo{volume}{29}}, \bibinfo{pages}{240}
  (\bibinfo{year}{1946}).

\bibitem[{\citenamefont{McKenna et~al.}(1995)\citenamefont{McKenna, Leterrier,
  and Schultheisz}}]{mck95}
\bibinfo{author}{\bibfnamefont{G.~B.} \bibnamefont{McKenna}},
  \bibinfo{author}{\bibfnamefont{Y.}~\bibnamefont{Leterrier}},
  \bibnamefont{and} \bibinfo{author}{\bibfnamefont{C.~R.}
  \bibnamefont{Schultheisz}}, \bibinfo{journal}{Polym. Eng. Sci.}
  \textbf{\bibinfo{volume}{35}}, \bibinfo{pages}{403} (\bibinfo{year}{1995}).

\bibitem[{\citenamefont{Mauro et~al.}(2009)\citenamefont{Mauro, Loucks, and
  Gupta}}]{mau09a}
\bibinfo{author}{\bibfnamefont{J.~C.} \bibnamefont{Mauro}},
  \bibinfo{author}{\bibfnamefont{R.~J.} \bibnamefont{Loucks}},
  \bibnamefont{and} \bibinfo{author}{\bibfnamefont{P.~K.} \bibnamefont{Gupta}},
  \bibinfo{journal}{J. Am. Ceram. Soc.} \textbf{\bibinfo{volume}{92}},
  \bibinfo{pages}{75} (\bibinfo{year}{2009}).

\bibitem[{\citenamefont{Ritland}(1956)}]{rit56}
\bibinfo{author}{\bibfnamefont{H.~N.} \bibnamefont{Ritland}},
  \bibinfo{journal}{J. Am. Ceram. Soc.} \textbf{\bibinfo{volume}{39}},
  \bibinfo{pages}{403} (\bibinfo{year}{1956}).

\bibitem[{\citenamefont{Hopkins}(1958)}]{hop58}
\bibinfo{author}{\bibfnamefont{I.~L.} \bibnamefont{Hopkins}},
  \bibinfo{journal}{J. Polym. Sci.} \textbf{\bibinfo{volume}{28}},
  \bibinfo{pages}{631} (\bibinfo{year}{1958}), ISSN \bibinfo{issn}{1542-6238}.

\bibitem[{\citenamefont{Morland and Lee}(1960)}]{mor60}
\bibinfo{author}{\bibfnamefont{L.~W.} \bibnamefont{Morland}} \bibnamefont{and}
  \bibinfo{author}{\bibfnamefont{E.~H.} \bibnamefont{Lee}},
  \bibinfo{journal}{Trans. Soc. Rheol.} \textbf{\bibinfo{volume}{4}},
  \bibinfo{pages}{233} (\bibinfo{year}{1960}).

\bibitem[{\citenamefont{Lee and Rogers}(1963)}]{lee63}
\bibinfo{author}{\bibfnamefont{E.~H.} \bibnamefont{Lee}} \bibnamefont{and}
  \bibinfo{author}{\bibfnamefont{T.~G.} \bibnamefont{Rogers}},
  \bibinfo{journal}{J. Appl. Mech.} \textbf{\bibinfo{volume}{30}},
  \bibinfo{pages}{127} (\bibinfo{year}{1963}).

\bibitem[{\citenamefont{McKenna}(1994)}]{mck94}
\bibinfo{author}{\bibfnamefont{G.~B.} \bibnamefont{McKenna}},
  \bibinfo{journal}{J. Res. Natl. Inst. Stand. Technol.}
  \textbf{\bibinfo{volume}{99}}, \bibinfo{pages}{169} (\bibinfo{year}{1994}).

\bibitem[{\citenamefont{Hecksher et~al.}(2010)\citenamefont{Hecksher, Olsen,
  Niss, and Dyre}}]{hec10}
\bibinfo{author}{\bibfnamefont{T.}~\bibnamefont{Hecksher}},
  \bibinfo{author}{\bibfnamefont{N.~B.} \bibnamefont{Olsen}},
  \bibinfo{author}{\bibfnamefont{K.}~\bibnamefont{Niss}}, \bibnamefont{and}
  \bibinfo{author}{\bibfnamefont{J.~C.} \bibnamefont{Dyre}},
  \bibinfo{journal}{J. Chem. Phys.} \textbf{\bibinfo{volume}{133}},
  \bibinfo{pages}{174514} (\bibinfo{year}{2010}).

\bibitem[{\citenamefont{Wehn et~al.}(2007)\citenamefont{Wehn, Lunkenheimer, and
  Lo}}]{weh07}
\bibinfo{author}{\bibfnamefont{R.}~\bibnamefont{Wehn}},
  \bibinfo{author}{\bibfnamefont{P.}~\bibnamefont{Lunkenheimer}},
  \bibnamefont{and} \bibinfo{author}{\bibfnamefont{A.}~\bibnamefont{Loidl}},
  \bibinfo{journal}{J. Non-Cryst. Solids} \textbf{\bibinfo{volume}{353}},
  \bibinfo{pages}{3862} (\bibinfo{year}{2007}).

\bibitem[{\citenamefont{Dyre and Olsen}(2003)}]{dyr03}
\bibinfo{author}{\bibfnamefont{J.~C.} \bibnamefont{Dyre}} \bibnamefont{and}
  \bibinfo{author}{\bibfnamefont{N.~B.} \bibnamefont{Olsen}},
  \bibinfo{journal}{Phys. Rev. Lett.} \textbf{\bibinfo{volume}{91}},
  \bibinfo{pages}{155703} (\bibinfo{year}{2003}).

\bibitem[{\citenamefont{Igarashi
  et~al.}(2008{\natexlab{a}})\citenamefont{Igarashi, Christensen, Larsen,
  Olsen, Pedersen, Rasmussen, and Dyre}}]{iga08a}
\bibinfo{author}{\bibfnamefont{B.}~\bibnamefont{Igarashi}},
  \bibinfo{author}{\bibfnamefont{T.}~\bibnamefont{Christensen}},
  \bibinfo{author}{\bibfnamefont{E.~H.} \bibnamefont{Larsen}},
  \bibinfo{author}{\bibfnamefont{N.~B.} \bibnamefont{Olsen}},
  \bibinfo{author}{\bibfnamefont{I.~H.} \bibnamefont{Pedersen}},
  \bibinfo{author}{\bibfnamefont{T.}~\bibnamefont{Rasmussen}},
  \bibnamefont{and} \bibinfo{author}{\bibfnamefont{J.~C.} \bibnamefont{Dyre}},
  \bibinfo{journal}{Rev. Sci. Instrum.} \textbf{\bibinfo{volume}{79}},
  \bibinfo{eid}{045105} (\bibinfo{year}{2008}{\natexlab{a}}).

\bibitem[{\citenamefont{Igarashi
  et~al.}(2008{\natexlab{b}})\citenamefont{Igarashi, Christensen, Larsen,
  Olsen, Pedersen, Rasmussen, and Dyre}}]{iga08b}
\bibinfo{author}{\bibfnamefont{B.}~\bibnamefont{Igarashi}},
  \bibinfo{author}{\bibfnamefont{T.}~\bibnamefont{Christensen}},
  \bibinfo{author}{\bibfnamefont{E.~H.} \bibnamefont{Larsen}},
  \bibinfo{author}{\bibfnamefont{N.~B.} \bibnamefont{Olsen}},
  \bibinfo{author}{\bibfnamefont{I.~H.} \bibnamefont{Pedersen}},
  \bibinfo{author}{\bibfnamefont{T.}~\bibnamefont{Rasmussen}},
  \bibnamefont{and} \bibinfo{author}{\bibfnamefont{J.~C.} \bibnamefont{Dyre}},
  \bibinfo{journal}{Rev. Sci. Instrum.} \textbf{\bibinfo{volume}{79}},
  \bibinfo{eid}{045106} (\bibinfo{year}{2008}{\natexlab{b}}).

\bibitem[{\citenamefont{Christensen and Olsen}(1995)}]{chr95}
\bibinfo{author}{\bibfnamefont{T.}~\bibnamefont{Christensen}} \bibnamefont{and}
  \bibinfo{author}{\bibfnamefont{N.~B.} \bibnamefont{Olsen}},
  \bibinfo{journal}{Rev. Sci. Instrum..} \textbf{\bibinfo{volume}{66}},
  \bibinfo{pages}{5019} (\bibinfo{year}{1995}).

\bibitem[{\citenamefont{Ellegaard et~al.}(2007)\citenamefont{Ellegaard,
  Christensen, Christiansen, Olsen, Pedersen, Schr{\o}der, and Dyre}}]{ell07}
\bibinfo{author}{\bibfnamefont{N.~L.} \bibnamefont{Ellegaard}},
  \bibinfo{author}{\bibfnamefont{T.}~\bibnamefont{Christensen}},
  \bibinfo{author}{\bibfnamefont{P.~V.} \bibnamefont{Christiansen}},
  \bibinfo{author}{\bibfnamefont{N.~B.} \bibnamefont{Olsen}},
  \bibinfo{author}{\bibfnamefont{U.~R.} \bibnamefont{Pedersen}},
  \bibinfo{author}{\bibfnamefont{T.~B.} \bibnamefont{Schr{\o}der}},
  \bibnamefont{and} \bibinfo{author}{\bibfnamefont{J.~C.} \bibnamefont{Dyre}},
  \bibinfo{journal}{J. Chem. Phys.} \textbf{\bibinfo{volume}{126}},
  \bibinfo{pages}{074502} (\bibinfo{year}{2007}).

\bibitem[{\citenamefont{Petrie}(1972)}]{pet72}
\bibinfo{author}{\bibfnamefont{S.~E.~B.} \bibnamefont{Petrie}},
  \bibinfo{journal}{J. Polym. Sci. A-2: Polymer Physics}
  \textbf{\bibinfo{volume}{10}}, \bibinfo{pages}{1255} (\bibinfo{year}{1972}).

\bibitem[{\citenamefont{Kolvin and Bouchbinder}(2012)}]{kol12}
\bibinfo{author}{\bibfnamefont{I.}~\bibnamefont{Kolvin}} \bibnamefont{and}
  \bibinfo{author}{\bibfnamefont{E.}~\bibnamefont{Bouchbinder}},
  \bibinfo{journal}{Phys. Rev. E} \textbf{\bibinfo{volume}{86}},
  \bibinfo{pages}{010501} (\bibinfo{year}{2012}).

\bibitem[{\citenamefont{Richert et~al.}(2013)\citenamefont{Richert,
  Lunkenheimer, Kastner, and Loidl}}]{ric13}
\bibinfo{author}{\bibfnamefont{R.}~\bibnamefont{Richert}},
  \bibinfo{author}{\bibfnamefont{P.}~\bibnamefont{Lunkenheimer}},
  \bibinfo{author}{\bibfnamefont{S.}~\bibnamefont{Kastner}}, \bibnamefont{and}
  \bibinfo{author}{\bibfnamefont{A.}~\bibnamefont{Loidl}}, \bibinfo{journal}{J.
  Phys. Chem. B} \textbf{\bibinfo{volume}{117}}, \bibinfo{pages}{12689}
  (\bibinfo{year}{2013}).

\end{thebibliography}
\end{document}